\documentclass[aps,prl,groupedaddress,twocolumn,superscriptaddress,amsfonts,amssymb,amsmath,showpacs,floatfix,preprintnumbers]{revtex4-2}

\usepackage{dcolumn}
\usepackage{bm}
\usepackage{multirow}
\usepackage{verbatim}
\usepackage{graphicx}
\usepackage{ulem}
\usepackage{color}
\usepackage{placeins}
\usepackage{xcolor}
\usepackage{hyperref}
\usepackage{slashed}
\usepackage{cleveref}

\begin{document}


\title{Comment on ``LaMET's Asymptotic Extrapolation vs. Inverse Problem''}

\author{Herv{\'e} Dutrieux}
\email[e-mail: ]{herve.dutrieux@cpt.univ-mrs.fr}
\affiliation{Aix Marseille Univ, Universit\'e de Toulon, CNRS, CPT, Marseille, France.}
\author{Joe Karpie}
\email[e-mail: ]{jkarpie@jlab.org}
\affiliation{Thomas Jefferson National Accelerator Facility, Newport News, VA 23606, U.S.A.}
\author{Christopher J.~Monahan}
\email[e-mail: ]{cjmonahan@coloradocollege.edu}
\affiliation{Department of Physics, Colorado College, Colorado Springs, CO 80903, U.S.A.}
\author{Kostas Orginos}
\email[e-mail: ]{kostas@jlab.org}
\affiliation{Physics Department, William \& Mary, Williamsburg, VA 23187, U.S.A.} 
\affiliation{Thomas Jefferson National Accelerator Facility, Newport News, VA 23606, U.S.A.}
\author{Anatoly Radyushkin}
\email[e-mail: ]{radyush@jlab.org}
\affiliation{Old Dominion University, Norfolk, Virginia 23529 U.S.A.}\affiliation{Thomas Jefferson National Accelerator Facility, Newport News, VA 23606, U.S.A.}
\author{David Richards}
\email[e-mail: ]{dgr@jlab.org}
\affiliation{Thomas Jefferson National Accelerator Facility, Newport News, VA 23606, U.S.A.}
\author{Savvas Zafeiropoulos}
\email[e-mail: ]{savvas.zafeiropoulos@cpt.univ-mrs.fr }
\affiliation{Aix Marseille Univ, Universit\'e de Toulon, CNRS, CPT, Marseille, France.}

\preprint{JLAB-THY-25-4392}

\begin{abstract}
In arXiv:2504.17706~\cite{Dutrieux:2025jed} we criticized the excessive model-dependence introduced by rigid few-parameter fits to extrapolate lattice data in the large momentum effective theory (LaMET) when the data are noisy and lose signal before an exponential asymptotic behavior of the space-like correlators is established. In reaction, arXiv:2505.14619~\cite{Chen:2025cxr} claims that even when the data is of poor quality, rigid parametrizations are better than attempts at representing the uncertainty using what they call ``inverse problem methods''. We clarify the fundamental differences in our perspectives regarding how to meaningfully handle noisy lattice matrix elements, especially when they exhibit a strong sensitivity to the choice of regularization in the inverse problem. We additionally correct misunderstandings of Ref.~\cite{Chen:2025cxr} on our message and methods. 
\end{abstract}

\maketitle

\section{What is at stake?}

In the framework of large momentum effective theory (LaMET) \cite{Ji:2013dva, Ji:2014gla}, computing the parton distribution function (PDF) requires solving an inverse problem: constructing a continuous function, known as the quasi-PDF, from a limited set of noisy computations of Fourier harmonics. We demonstrated in Ref.~\cite{Dutrieux:2025jed} that due to the quickly increasing noise of larger harmonics in many current lattice studies, and even including physical assumptions on the asymptotic behavior of the missing harmonics, this inverse problem suffers from a strong sensitivity to the choice of regularization technique. We have, in fact, demonstrated that the exact asymptotic behavior represents a minor source of systematic uncertainty on the calculation compared to the choice of regularization technique. Therefore, we stressed the need for a more comprehensive study of uncertainty in the determination of PDFs from noisy lattice datasets. 

The authors of Ref.~\cite{Chen:2025cxr} restated their strong preference for rigid parametric extrapolation fits over more sophisticated uncertainty treatment, even in the case of noisy limited datasets. They fear that our proposed treatment of the inverse problem could result in ``{\it unnecessarily conservative errors}''. Their main arguments can be summarized as follows:
\begin{enumerate}
\item Constructing a continuous function from a limited set of noisy Fourier harmonics appears to them, not as an inverse problem, but rather as a forward problem with an extrapolation issue.
\item A physical analysis demonstrates that the missing harmonics decay exponentially, so they hold that using a simple parametric form of exponential decay is a controlled extrapolation method although in their own words it ``{\it might be perceived as a model}''. We note that for many LaMET datasets, data with good signal for the renormalized matrix elements barely extend to one exponential decay length for the typically-quoted value of the decay length at physical quark masses and useful hadron momenta. 
\item They introduce a number of rules on the basis of which they declare our proposed ``{\it inverse problem methods}'' unphysical, whereas they consider their parametric fit extrapolations generally physical. 
\item They attempt to apply our method to their own dataset and conclude that it obviously produces unphysical distortions.
\item They restate that the exponential decay is truly fundamental to controlled uncertainty in the moderate to large $x$ region in contrast to our statement that the assumed smoothness of PDFs is the meaningful physical regulator.
\end{enumerate}

\begin{figure}
    \centering
    \includegraphics[width=0.9\linewidth]{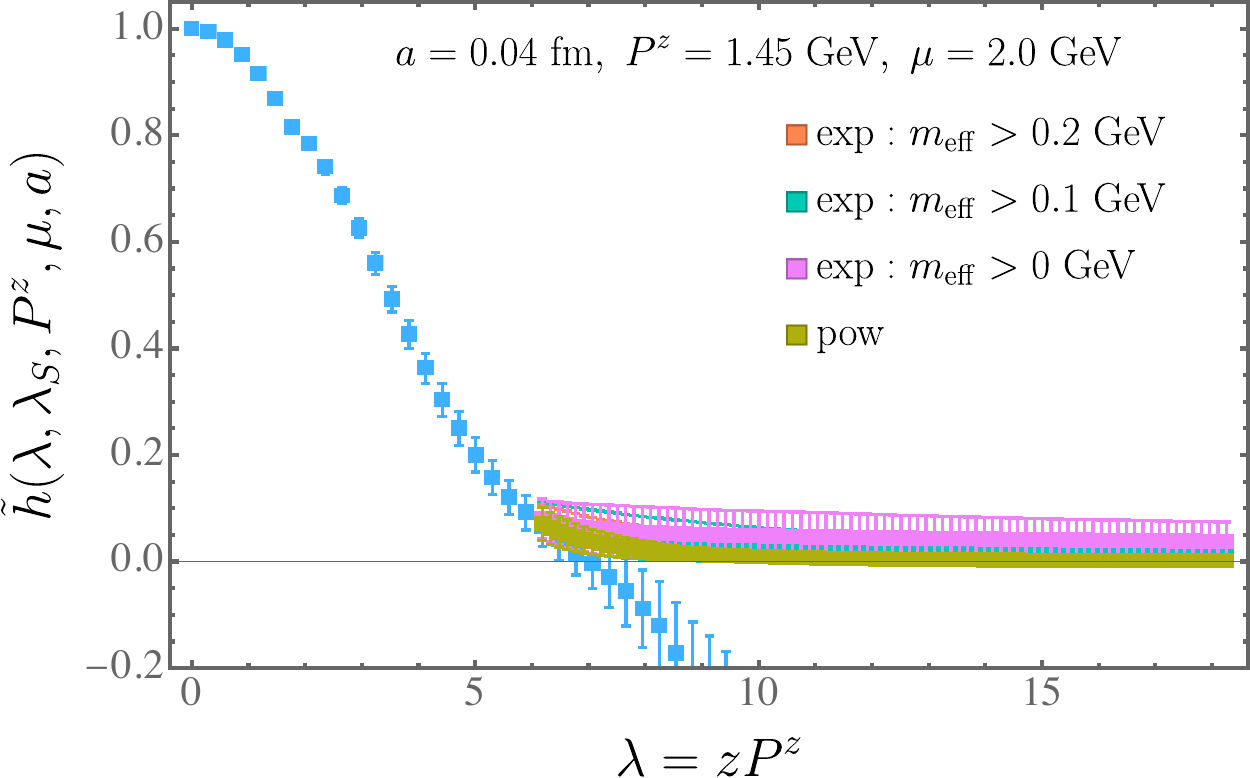}
    \caption{The set of Fourier data for the pion valence quark PDF and its proposed parametric fit extrapolations from Ref.~\cite{Gao:2021dbh} (pion mass 300 MeV).
    }
    \label{fig:fig1}
\end{figure}

\begin{figure}
    \centering
    \includegraphics[width=0.9\linewidth]{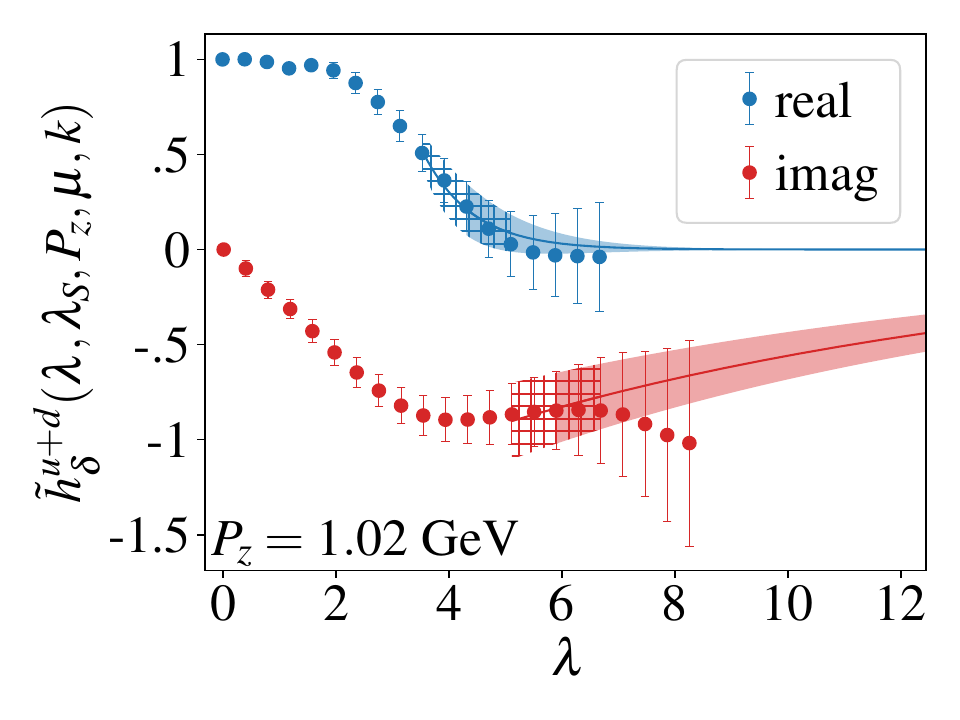}
    \caption{The set of Fourier data (real and imaginary parts) for the $u+d$ transversity PDF of the proton at physical quark masses and its proposed extrapolation from Ref. \cite{Gao:2023ktu}.}
    \label{fig:fig2}
\end{figure}

We will show in the following that, in our opinion, none of these points are valid. But before we do so, we believe that it is important for a reader who may not be entirely familiar with the problem at hand to observe for themselves what exactly constitutes the rigid parametric extrapolation defended by Ref.~\cite{Chen:2025cxr}. This is what motivated us to write our study in Ref.~\cite{Dutrieux:2025jed} in the first place. We reproduce in Figs.~\ref{fig:fig1} and \ref{fig:fig2} examples of parametric fit extrapolations that some of the authors of Ref.~\cite{Chen:2025cxr} have produced in recent years.

In Fig.~\ref{fig:fig1}, extracted from Ref.~\cite{Gao:2021dbh}, the lattice matrix elements drift significantly below zero beyond a Fourier harmonic of $\lambda = 7$. The preferred method of parametric extrapolation of the authors of Refs.~\cite{Gao:2021dbh,Chen:2025cxr} consists of ignoring this behavior and replacing the data starting from the region where they approach zero with their modeling of a decaying tail. Ref.~\cite{Gao:2021dbh} studies several parametric extrapolations, including an exponential decay with surprisingly light decay mass or ``{\it pow}'' which uses an unphysical power-law decay instead of an exponential one. It is observed that all parametric extrapolations give seemingly similar $x$-reconstructions, except at small $x$ where it is understood that the reconstruction is unreliable in any case. This observation supports our claim that the asymptotic behavior matters little to the reconstruction at moderate values of $x$.

In Fig.~\ref{fig:fig2}, extracted from Ref.~\cite{Gao:2023ktu}, the extrapolation in the imaginary part starts when the signal is still $4 \sigma$ away from zero. There is an obvious large model dependence not just in the standard deviation and correlations of the extrapolation, but also in its central value. One should note that, by the standards set in Ref.~\cite{Chen:2025cxr}, this is not a physically meaningful extrapolation since it starts when the lattice data still presents a strong non-vanishing signal. The authors of Ref.~\cite{Chen:2025cxr} do not provide clear guidance on what to do in that case, in contrast to our proposed method. 

For the real part, the proposed extrapolation decreases in uncertainty at a remarkably fast rate. Closely examining the figure allows us to determine that the effective mass controlling the exponential decay is of the order of 1 GeV. Ref.~\cite{Gao:2023ktu} does not attempt to explain why such a large value is physically meaningful when Ref.~\cite{Chen:2025cxr} mentions that one expects an effective mass controlling the decay of the order of 0.2 to 0.3 GeV.

Our explanation for this very large discrepancy is simple: there is no conclusive sign of exponential decay in the noisy fitted dataset. Since the data actually slightly inch towards the negative domain, performing an exponential parametric fit to a model that cannot change sign results in an unphysically fast exponential decay. One notes that there is no conclusive sign of exponential decay in any of the datasets that we have shown here, as is the case in a significant number of lattice studies. Let us therefore restate an opening statement of our previous work \cite{Dutrieux:2025jed}: when noisy data loses signal before or around one expected decay length, it is difficult to believe that a simple parametric fit attempting to capture this decay will produce meaningful results.

Our work~\cite{Dutrieux:2025jed} had a deeper objective than making such observations. We laid the groundwork for a possible method to improve the uncertainty quantification in the reconstruction of PDFs from noisy lattice data and we restate aspects of our approach later. 

Our respective positions on uncertainty quantification are nicely summarized by the authors of Ref.~\cite{Chen:2025cxr}: ``{\it When data in the sub-asymptotic region is noisy and the number of fit parameters exceeds the number of reliable data points, the fits are unstable. [...] In this case, one should either generate more and better data or limit the number of parameters in the fitting and estimate the errors based on the relevant physics conditions. [...] On the other hand, it has been suggested in Ref.~\cite{Dutrieux:2025jed} that in LaMET analysis, if the lattice data have large errors in the sub-asymptotic domain, one should treat the extrapolation as an IP [inverse problem]}''. 

The strategy of the authors of Ref.~\cite{Chen:2025cxr} is clear: when the data is too noisy, they increase the bias in the modeling (``{\it limit the number of parameters in the fitting}'') to maintain the appearance of good control of the systematic uncertainty. They state that their extrapolation approach provides a more ``accurate'' result, while we suggest that they only demonstrate it is more ``precise''. In contrast, we argue that the less informative the data, the more ill-defined the inverse problem is, and the more careful the uncertainty assessment must be. What is really at stake is therefore whether the uncertainty obtained as a result of the rigid few-parameter fits advocated by the authors of Ref.~\cite{Chen:2025cxr} is trustworthy. 

Finally, on top of several misunderstandings in our proposal that we correct below, two points must be stressed. We did not conclude in Ref.~\cite{Dutrieux:2025jed} that it is not possible to quantify or control the uncertainties in the LaMET inverse problem as the authors of Ref.~\cite{Chen:2025cxr} repeatedly write. We merely point out that for noisy datasets, their technique underestimates significantly the error in the reconstruction and more advanced uncertainty quantification is necessary. We also took care in Ref.~\cite{Dutrieux:2025jed} to mention several times that the problem we address concerns datasets where the noise is large in the range $\lambda \sim 5$ -- $15$. For studies where the uncertainties are minimal in this range, the concerns linked to the Fourier transform at moderate $x$ are indeed minimal compared to many other systematics. The dataset used by Ref.~\cite{Chen:2025cxr} for their example exhibits a significantly lower level of noise than those we have mentioned above or the one we used in our own study in Ref.~\cite{Dutrieux:2025jed}, so it does not come as a surprise that the uncertainty propagation is under better control.

Ref.~\cite{Xiong:2025obq}, co-authored by some of the authors of Ref.~\cite{Chen:2025cxr}, appeared when we were about to submit this work to the arXiv. They confirm that there is an inverse problem in the LaMET formalism and that it has a ``{\it moderately tractable}'' ill-posedness. Therefore, they seem to agree on most relevant points with our own analysis and disagree with Ref.~\cite{Chen:2025cxr} which holds that there is no inverse problem in LAMET and that one should not use ``{\it inverse problem methods}'' to study uncertainty propagation. However, Ref.~\cite{Xiong:2025obq} uses a simple diagonal Tikhonov regulator, a special case of the Gaussian Process Regression we employ in Ref.~\cite{Dutrieux:2025jed}, which we consider less suited than regulators that implement a correlated structure. A notable advantage of regulators with correlation is the reduction of uncertainty tightening effect, as we explored in Ref.~\cite{Dutrieux:2024rem}. Furthermore, in contrast to our work, Ref.~\cite{Xiong:2025obq} does not seek to enforce physics-driven conditions on the asymptotic behavior of the space-like correlator.

\section{Detailed answer to arXiv:2505.14619}

\subsection{Differences in definitions}

The comment of Ref.~\cite{Chen:2025cxr} on our work focuses first on a point of definition.
The authors of Ref.~\cite{Chen:2025cxr} do not agree that constructing a continuous function from a limited set of Fourier harmonics constitutes an inverse problem in the context of the LaMET formalism. To fix notations, $f(y, P_z)$ is the quasi-PDF, a continuous function of the $y$ parameter at fixed hadron momentum $P_z$. It is constructed from a limited, discrete, noisy set of hadronic matrix elements $h(z, P_z)$ computed on the lattice, where $z$ is the space-like separation of the non-local field operator of which the matrix element is taken. If all $h(z, P_z)$ were known for all values of $z$, the relation between $f(y, P_z)$ and $h(z, P_z)$ would be given by the simple Fourier transform:
\begin{equation}
f(y, P_z) = P_z \int_{-\infty}^\infty \frac{dz}{2\pi} e^{i y P_z z} h(z, P_z)\,. \label{eq:invPb}
\end{equation}
However, since only a few discrete values of $h(z, P_z)$ for some $z$'s below $z_{\max}$ are known, and with quickly growing uncertainty as $z$ increases, one cannot directly apply the formula of Eq.~\eqref{eq:invPb}. The fact that a general expectation on the asymptotic behavior of the missing Fourier harmonics is known leads the authors of Ref. \cite{Chen:2025cxr} to qualify this as a ``{\it forward problem}'' with ``{\it essentially a matter of convergence and precision of the numerical Fourier integral, which are influenced by the discretization and the extrapolation in $z$-space}''. 

It seems to us likely that a number of problems that the authors of Ref. \cite{Chen:2025cxr} themselves recognize as inverse problems can be rebranded as ``forward problems'' with a matter of discretization and extrapolation of the limited data. Therefore, either problems commonly accepted as paradigm examples of inverse problems within hadronic physics, such as the extraction of PDFs from deep inelastic scattering (DIS) structure functions, are in fact ``forward problems" or one is forced to conclude that the determination of quasi-PDFs from lattice data is an inverse problem, on pain of inconsistency.
To put this more concretely, consider the phenomenological inverse problem of determining a PDF $q(x)$ from DIS structure functions $F(x_B)$. Although this is not its most typical formulation, it can be written, up to power corrections, as:
\begin{equation}
q(x, \mu^2) = \int_x^1 \frac{d x_B}{x_B} C\left(\frac{x}{x_B}, \mu^2, Q^2\right) F(x_B, Q^2)\,,
\end{equation}
where $C$ is known to a given order in perturbation theory. The problem is then quite similar to the one faced above: one expresses a continuous function of the $x$ parameter from a finite set of noisy measurements of the structure function $F(x_B, Q^2)$. Using such formulation, the main issue of the inverse problem of PDF determination becomes an interpolation problem of the noisy discrete data in the region where data exist, and an extrapolation problem in the large/small $x_B$ regions where no data are available. For instance, since it is well-known that structure functions decrease regularly to 0 when $x_B$ tends to 1, one can propose various extrapolation strategies, and qualify this well-known inverse problem as a forward problem with a matter of convergence and precision of the discretization and extrapolation.

Furthermore, a central aspect of our study in Ref.~\cite{Dutrieux:2025jed}, which will be revisited at the end of this document, is that the exact asymptotic behavior of the space-like lattice matrix elements matters little in the inverse problem in either LaMET or pseudo-PDFs in the moderate $x$ region with which we are concerned. Therefore, it simply cannot be that the precise expectation of this asymptotic behavior changes fundamentally the characterization of the problem at hand. Instead, as we have demonstrated in Ref. \cite{Dutrieux:2025jed}, a more important regulator can be the prior expectation on the smoothness (or more precisely the correlation structure) of the $x$-dependence of the parton distribution. For rigid parametric fits, the regulator is the rigidity of the parametric form rather than its asymptotic behavior, explaining why fits that do not present the correct asymptotic behavior still give relatively similar uncertainty assessments at moderate $x$, as in Fig.~\ref{fig:fig1}.

The tell-tale sign of the existence of an ill-posed inverse problem is not whether or not one can formally revert the integral relationship between the continuous quantity of interest and the discrete noisy dataset, nor whether one has general or rigorous expectations on the behavior of the missing data. It is whether the final reconstruction exhibits a strong dependence on the choice of regularization of the inverse problem. The authors of Ref.~\cite{Chen:2025cxr} recommend studying only one method of regularization: replacing missing data, or data that they consider unreliable, with a simple rigid parametric fit. Since they do not recognize the existence of an inverse problem, they do not recognize that this procedure constitutes a model to regularize the inverse problem. That the small parametric variations within their class of models do not produce appreciable variation of uncertainty is not evidence that there is no inverse problem, but rather a testimony to their limited exploration of physically meaningful regulators.

This leads to the second main difference in definition, namely the opposition established by the authors of Ref.~\cite{Chen:2025cxr} between their parametric fit extrapolation procedure and what they call ``{\it inverse problem methods}''. Parametric fits to complete missing data are one of the most basic strategies to regularize an inverse problem and have been used successfully for decades to handle inverse problems in QCD phenomenology. They are a perfectly valid approach, provided one introduces adequate flexibility in the parametrization and the data is of high enough quality to support this flexibility. 
We do not believe, however, that many current lattice datasets are of sufficient quality to enable quantitative uncertainty estimates to be drawn from this approach. A more elaborate treatment of uncertainty is therefore necessary, some aspects of which we developed in Ref.~\cite{Dutrieux:2025jed}.

The authors of Ref.~\cite{Chen:2025cxr} believe that our discussion in Ref.~\cite{Dutrieux:2025jed} raises concerns ``{\it whether it becomes an inverse
problem where the relevant uncertainties cannot be properly quantified}''. Acknowledging the existence of an ill-posed inverse problem is not believing that uncertainties cannot be properly quantified, as decades of studies in QCD phenomenology demonstrate. Ref.~\cite{Chen:2025cxr} compares the LaMET inverse problem to the extraction of hadronic energies and matrix elements in lattice QCD from a series of exponentials. While they are right that the expression ``\textit{inverse problem}'' is not typically used to qualify those analyses, it obviously applies. For instance, there is a marked dependence of results on the truncation of asymptotic models, a hallmark of regulating an ill-posed inverse problem. In lattice QCD this specific issue has been studied for decades with premature truncation of the asymptotic series in the sub-asymptotic regime causing issues, typically called excited state contamination, which can lead to results statistically incompatible with the true answer. The application of this specific asymptotic model is now very well understood and yet modern Bayesian techniques are still being introduced for resolving the inverse problem while controlling bias from truncation of that model~\cite{Jay:2020jkz,Neil:2023pgt}. An identical problem can arise in LaMET if a truncated model is applied in an inappropriate regime and the same Bayesian tools can be used to create a more robust uncertainty quantification from the results of different truncations of models~\cite{Karpie:2021pap}.
 
\subsection{Our proposal to study uncertainty}

Let us recap briefly the method in Ref.~\cite{Dutrieux:2025jed} to enforce a physical asymptotic decay without using a simple rigid model. On top of the computed dataset, which we use in its entirety without performing any truncation, we add two Bayesian priors. The first is a prior on the mean and covariance of the reconstruction in $x$-space as we presented before in Ref.~\cite{Dutrieux:2024rem}. This is by far the most important prior, which encodes our expectations in terms of smoothness in $x$ and degree of correlation between various regions in $x$. Making different, yet physically sensible choices on this prior produces significant differences in the final uncertainty assessment. This arises naturally from the ill-posedness of the inverse problem when the lattice dataset is poorly constraining.

\begin{figure}
    \centering
\includegraphics[width=0.95\linewidth]{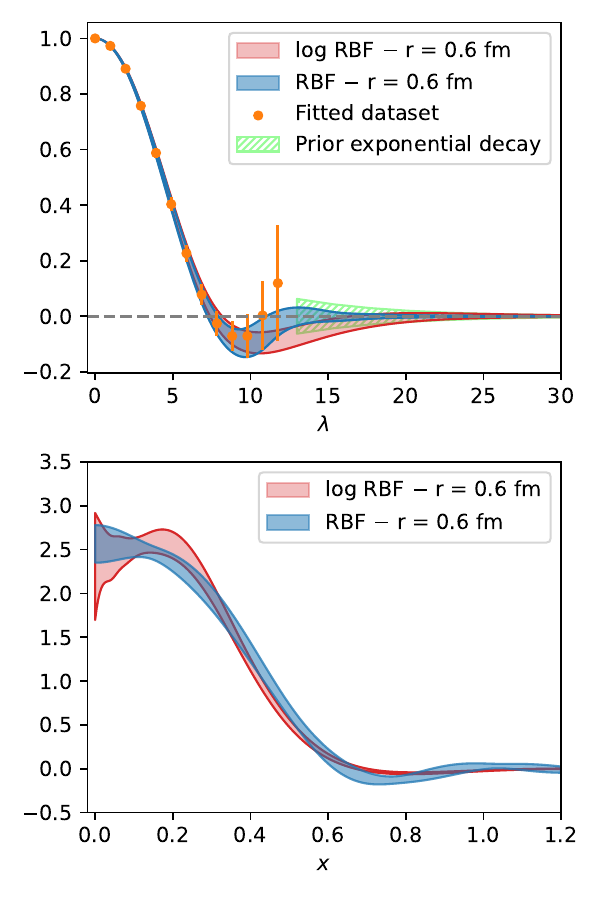}

\includegraphics[width=0.95\linewidth]{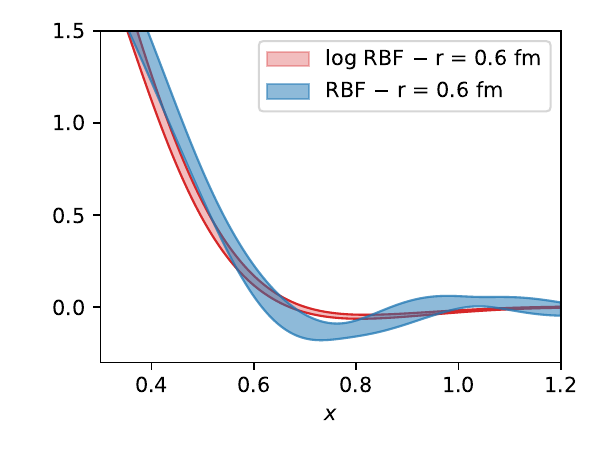}

    \caption{(top) Two solutions of the inverse problem in Fourier space where the exponential decay of the missing harmonics is enforced by the same green prior, corresponding to a decay length of 0.6 fm or a mass governing the decay of 0.33 GeV. The two solutions differ by their regularization thanks to a different prior kernel in $x$-space. (middle) The associated $x$-reconstructions with a zoom on the large $x$ region (bottom) to appreciate the very different uncertainty assessment produced by the two regularizations of the inverse problem.}
    \label{fig:nfig1}
\end{figure}

An additional Bayesian prior is enforced in the Fourier space to imprint the desired exponential decay of space-like matrix elements at large $z$ separation. A representation of the situation is given in Fig.~\ref{fig:nfig1} where one can appreciate the variance (unfortunately not the correlation length which is also an integral aspect of the prior setting) of a Bayesian prior enforcing an asymptotic exponential decay through the green band \footnote{Unlike what the authors of Ref.~\cite{Chen:2025cxr} seem to have understood, we do not enforce the exponential decay prior on the entire range in $\lambda$, but only when $\lambda$ is larger than the largest value in the dataset. However, it should not make a large difference as the uncertainty of the prior would become considerably larger than that of the dataset if it was also imposed at small $\lambda$.}.

The main objection of Ref.~\cite{Chen:2025cxr} against our proposal is that it would be unphysical by the three ``{\it physical conditions}'' they have introduced in Ref.~\cite{Chen:2025cxr}:
\begin{enumerate}
\item ``The lattice data should be truncated at a point within the sub-asymptotic region where they have decreased smoothly to near zero.''
\item ``The extrapolation model must preserve the correct asymptotic behavior of the correlation function without introducing unphysical distortions.''
\item ``The extrapolated results should continue to decrease in amplitude even with the presence of oscillations, consistent with asymptotic decay.''
\end{enumerate}

The first rule stating that the lattice data should be truncated is arbitrary. Truncating the data and replacing the missing data with some extrapolation is just one way to regularize the inverse problem. Our method proposes to construct an $x$-dependence which is in excellent statistical agreement with the lattice dataset and includes no truncation. We consider this a superior method as it does not lead to simply neglecting parts of the computed dataset as in the case of Fig.~\ref{fig:fig1}. If one believes that the lattice data is corrupted beyond a certain value of $z$ and truly needs to be ignored, then one should provide a convincing demonstration that this corrupting mechanism has not also biased the results at smaller values of $z$. We see no reason to perform a truncation unless compelling evidence constrains us to do it, and a mere negative central value of the lattice matrix elements, which was the truncation criteria in Refs.~\cite{Gao:2021dbh,Gao:2023ktu}, does not appear to us as a compelling reason.

The second rule is well-understood, and the Bayesian prior guarantees its application in a less model-dependent fashion than a simple parametric fit. To correctly implement our method, one must set priors differently for different datasets \footnote{It may be natural to question the use of the term ``prior'' if our priors depend in fact on the dataset, and ultimately on the final reconstruction, also known as ``posterior''. Our priors implement real elements of prior expectations independent of the dataset, such as the structure of correlations in $x$-space or the desired exponential decay rate in Fourier space. They also include elements that we have no prior knowledge about, such as appropriate scaling factors of the uncertainty, which depend on the datasets on top of the expectations of the physicist. In the end, our priors are really designed to enforce a desired pre-conceived physical behavior on the posterior. Therefore, they do represent a form of prior knowledge, and follow the ordinary mathematics of Bayesian priors. We keep therefore the standard terminology, as does Ref.~\cite{Candido:2024hjt} whose prior hyperparameters are also set with respect to the fitted dataset.}. For instance, more precise datasets require tighter priors to avoid unwanted features arising from the fluctuations of the extrapolation within the wide prior. Appropriate setting of hyperparameters is therefore necessary to alleviate tensions with the second rule. Fig.~2 of Ref.~\cite{Chen:2025cxr} casts doubt on the use of our Bayesian prior by presenting clearly faulty reconstructions. However, they do not use our prescription to set hyperparameters, which depends dynamically on the dataset. Instead, they use either a very short correlation length which allows for wide oscillations of the final result (blue curve) or hyperparameters obtained for another dataset resulting in a very large uncertainty (red curve). What this merely demonstrates is that improper hyperparameter setting results in improper regularization of the inverse problem. One can easily compare Fig.~2 of Ref.~\cite{Chen:2025cxr} with Fig.~\ref{fig:nfig1} here, or Fig.~3 of Ref.~\cite{Dutrieux:2025jed} to see that we do not suffer from such unphysical issues because we take care to fix hyperparameters in a meaningful fashion. Even then, strong dependence on the regularization choice persists.

There are many ways one can choose to set hyperparameters, from a simple examination of the effect of various choices, which is the method we have chosen in Ref.~\cite{Dutrieux:2025jed}, to statistical procedures to set hyperparameters through evidence sampling (see \textit{e.g.} \cite{Candido:2024hjt}) or studying how rapidly the size of the data likelihood and prior change with the hyperparameters (see \textit{e.g.} \cite{Xiong:2025obq}). We do not pretend to have produced a fool-proof way to make this crucial choice. The authors of Ref.~\cite{Chen:2025cxr} seem to see in this under-determination a weakness of our method. Yet, it is the simple manifestation of the ill-posed nature of the inverse problem that they do not acknowledge: one needs to make choices, and different choices, although reasonable, lead to a very different picture of the final uncertainty. A valuable uncertainty quantification reflects the variability of the answer with respect to those choices. Deciding that only one method is appropriate leads to a reduced final uncertainty, but equally to a reduced usefulness and reliability of the physical result. Further studies of different choices within the Gaussian Process approach will be published soon in upcoming work.

In Ref.~\cite{Dutrieux:2024rem} the choice of hyperparameters was specifically designed to not underestimate the errors in closure tests. This highlights the primary differences between the approach suggested in Ref.~\cite{Dutrieux:2025jed} and Ref.~\cite{Chen:2025cxr}. The goal of Ref.~\cite{Dutrieux:2025jed} is to attempt to ensure a sufficiently conservative, yet accurate, uncertainty quantification that can encapsulate any physically reasonable features of the final PDF. The goal of Ref.~\cite{Chen:2025cxr} is to assume the dominance of limiting behavior, possibly missing genuine features, in order to obtain the most precise result possible.

As for the third rule, a more precise formulation would be useful. It appears to us that all our proposed reconstructions do in fact decrease in amplitude in the extrapolation region, in a way which is commensurate with the uncertainties of the dataset and generally physically reasonable. The third rule mentions plausible ``oscillations''. Such features are prominently observed for instance in the pion distribution amplitude~\cite{Baker:2024zcd}, but their presence in ordinary PDFs cannot be assessed in many calculations due to the poor data quality. Oscillations make it difficult to identify whether the small size of a matrix element is due to a zero crossing linked to an ongoing oscillation, or if the small size truly reflects the exponential decay of the quantity. This highlights again the arbitrariness of truncating the dataset to replace it with a rigid model on the basis that it has decreased to near zero. 

In short, we doubt that the three rules of Ref.~\cite{Chen:2025cxr} provide a clear and useful framework to handle the inverse problem at hand. Furthermore, we do not clearly see how reconstructions such as that of Fig.~\ref{fig:nfig1} violate meaningfully those rules. On the other hand, the method of parametric extrapolation of the authors of Ref.~\cite{Chen:2025cxr} suffers from plausible non-physical issues that we have already evoked in the introduction and re-summarize here:
\begin{itemize}
\item the physical parameters are unlikely to be well constrained by noisy data which ends around one exponential decay length, leading for instance to an abnormally fast decay of uncertainty as in the real part of Fig.~\ref{fig:fig2}
\item the parametric extrapolation may be in strong tension with the truncated dataset as in Fig.~\ref{fig:fig1}
\item the reconstruction may suffer from very large model-dependence at the level of its central value if the dataset does not extend sufficiently in Fourier space as in the imaginary part of Fig.~\ref{fig:fig2}
\end{itemize}
Beyond those plausible non-physical issues remains the more fundamental issue of underestimation of the uncertainty linked to the model-dependence. Such issues are largely avoided with our procedure. In particular, there may exist a tension between our reconstruction and the dataset, but then it will be understood in terms of a tension between the dataset and the enforced priors. This allows us to understand the roots and consequences of the tension in a more illuminating fashion than a simple arbitrary truncation threshold.

\subsection{Inverse problem and upper-bound on uncertainty}

To back their claim that there is no inverse problem, the authors of Ref.~\cite{Chen:2025cxr} use again the demonstration of a ``rigorous upper-bound'' that some of them have been associated with in Ref.~\cite{Gao:2021dbh}. To ground our discussion, we first summarize the reasoning they employ. The starting point for this derivation is the observation that, if the missing harmonics decay exponentially, they must not really contribute meaningfully in the Fourier transform beyond a few $N = {\cal O}(1)$ exponential decay lengths and we can as well assume that they vanish beyond this. Then one obtains a bound proportional to this value $N$. This means that depending on whether one considers that one, two, or more decay lengths are necessary to neglect the missing parts of the integral, one will obtain an equally varying bound. This is not really an upper bound either since parts of the quantity to bound have been outright ignored instead of quantified. Yet, we agree that there must be a value of $N$ of the order of a few units where this broad estimate is reasonable. 

For typically available lattice studies, by their own calculations, the authors of Ref.~\cite{Gao:2021dbh, Chen:2025cxr} produce an upper-bound on the absolute uncertainty for $x \sim 0.5$ of the order of 0.25, by assuming $N=1$ and an absolute uncertainty on the largest considered Fourier harmonic of 0.1. They notice that this corresponds approximately to the general envelope of uncertainty that we have highlighted ourselves in our various reconstructions of the dataset we used in Ref.~\cite{Dutrieux:2025jed}. Therefore, are we not all agreeing in the end? Not quite. We have never doubted that one can produce a reasonable estimate on the maximal uncertainty of the $x$-reconstruction, depending obviously on some broad characteristics of the fitted dataset. In fact, we are convinced that this maximal uncertainty estimate holds regardless of the exponential decay, based solely on the general expectation that (usual) parton distributions are smooth and therefore the contribution of Fourier harmonics much beyond the inverse of the characteristic length of their variations are suppressed and irrelevant. In fact, prior to the publication of this upper bound in Ref.~\cite{Gao:2021dbh}, closure tests of multiple non-parametric approaches in Refs.~\cite{Karpie:2019eiq,Alexandrou:2020tqq} reproduced PDFs within this bound without assuming exponential decay of the data.

Repeating our introductory statements, what is at stake here is whether the uncertainty obtained as a result of the rigid few-parameter fits advocated by Ref.~\cite{Chen:2025cxr} is trustworthy, not whether there exists a theoretical upper-bound on the uncertainty. It seems clear to us that an analysis of the real part in Fig.~\ref{fig:fig2} that is less model-dependent would result in considerably larger uncertainties than those quoted in Ref.~\cite{Gao:2023ktu}, given that the final uncertainties on the $x$-reconstructions of Ref.~\cite{Gao:2023ktu} are significantly smaller than those we obtained although our dataset extended further in Fourier space with significantly smaller errors. Therefore, if the authors of Ref.~\cite{Chen:2025cxr} do not wish to perform a serious exploration of the model-dependence involved in their extrapolations, critical readers may find it useful to add a blanket absolute uncertainty to their statistical errors based on their own ``upper-bound''. If, however, the colored bands shown in the publications are to be taken seriously, then more work needs to be done to properly quantify uncertainties.

\subsection{Point-by-point reconstruction in the LaMET formalism}

Ref.~\cite{Chen:2025cxr} states ``{\it \dots LaMET can provide a point-by-point calculation — rather than fitting — of the $x$-dependence of parton distributions, which resolves the inverse problem faced by the pseudo-PDF due to limited range of accessible leading-twist correlations.}'' The previous examples and our work in Ref.~\cite{Dutrieux:2025jed} demonstrate a strong sensitivity to the choice of regularization of the inverse problem, which affects globally the whole $x$-dependence and is at odds with the claim of point-by-point determination of Ref.~\cite{Chen:2025cxr}.

\begin{figure*}
    \centering
    \includegraphics[width=0.7\linewidth]{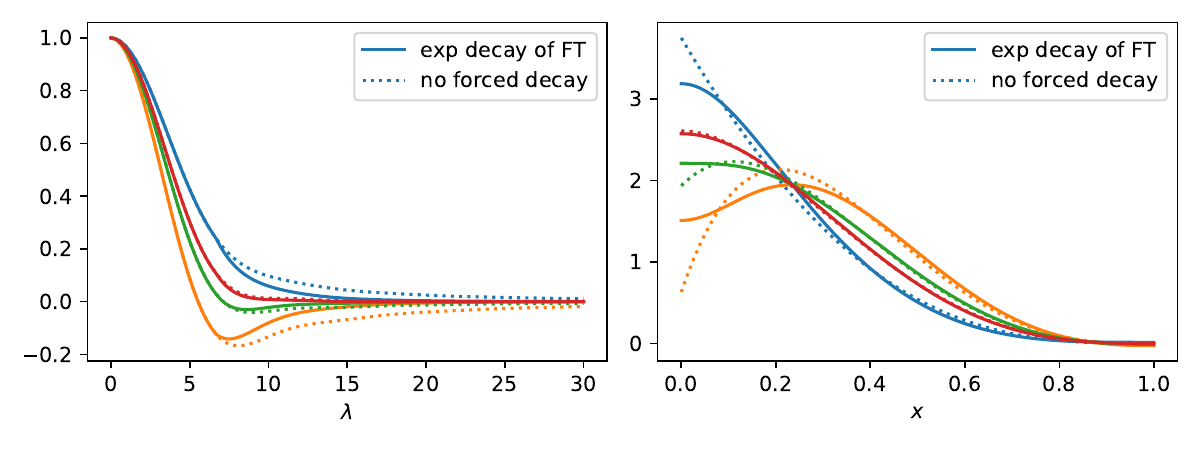}
    \includegraphics[width=0.7\linewidth]{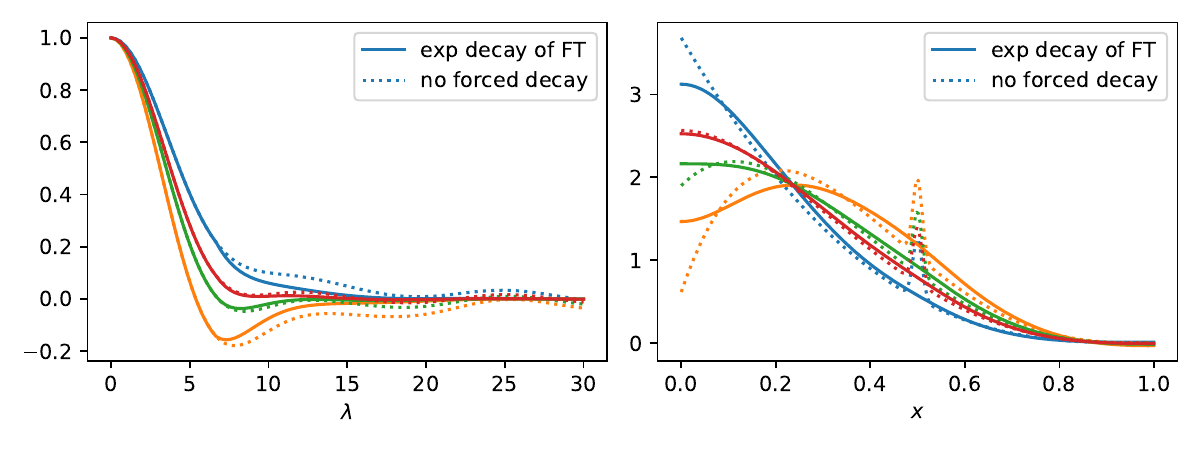}
    \caption{(top) The dotted lines represent the Fourier transform (left) and $x$-space (right) models obtained following Eq.~\eqref{eq:modelquasi}. The solid lines represent the same model with a physically relevant exponential reduction of large harmonics. The region $x > 0.3$ is barely impacted by the exponential decay. -- (bottom) We now add an unphysical bump at $x = 0.5$ (dotted lines). When the exponential decay is enforced (dotted lines), the bump is erased in the $x$-reconstruction.}
    \label{fig:force_decay}
\end{figure*}

To investigate the role of the exponential decay of missing Fourier harmonics again, on top of our previous explorations in Ref.~\cite{Dutrieux:2025jed}, we present in Fig. \ref{fig:force_decay} a number of models of quasi-PDF obtained by simply varying the coefficient $a$ in the expression
\begin{equation}
f(x) = a (1-x)^3 - (1-x)^4 - (1-x)^5\,.\label{eq:modelquasi}
\end{equation}
The Fourier harmonics $h(\lambda) = \int_0^1 \mathrm{d}x\,\cos(x \nu)f(x)$ of this function decrease generally as $1 / \lambda^2$ at large $\lambda$, so much more slowly than the expected exponential decay. They are represented by the dotted curves in the upper plots of Fig. \ref{fig:force_decay}. 

Then we enforce a typical exponential decay corresponding to a mass $m_{\textrm{eff}} = 0.3$ GeV and a hadron momentum $P_z = 2$ GeV. The decay is obtained by multiplying $h(\lambda)$ by $\exp(1 - \lambda m_{\textrm{eff}}/P_z)$ when $\lambda > P_z / m_{\textrm{eff}}$. The result is displayed in the top plots by the solid lines. The marked decrease of the amplitude of large Fourier harmonics is visible. In terms of $x$-reconstruction however, the effect of exponential reduction of the FT is barely noticeable in the range [0.3, 1]. Only for $x < 0.3$ does the implementation of exponential suppression result in a significant effect. This confirms once again a result we established in Ref.~\cite{Dutrieux:2025jed}, namely that the asymptotic behavior of the Fourier harmonics barely matters in the relevant range in $x$ under sensible assumptions on the smoothness of parton distributions \footnote{One could be tempted to believe again that, since one can modify the extrapolation region on a given model with little impact on the $x$-reconstruction, then there is no inverse problem at all. The reason would then not be that LaMET escapes the inverse problem thanks to its physical assumptions on large Fourier harmonics, but that there is no serious issue with not knowing the large Fourier harmonics at all. This is clearly not true since it would apply indiscriminately to all lattice formalisms, including the well-established inverse problem in pseudo-PDFs. The inverse problem lies in the fact that many real datasets have large uncertainties starting from $\lambda \approx 5 - 8$ and that one can present very different $x$-reconstructions that are statistically compatible with those datasets. The inverse problem is very improperly understood when looking only at the impact of small fluctuations on an error-less single model since it is a matter of stability of reconstruction \textit{within errors}. However, this allows us to isolate the impact, or rather lack of impact, of the precise asymptotic behavior.}.

Another useful observation is that the Fourier transform of the orange model takes markedly negative values, although it corresponds to a fully positive and seemingly reasonable model in $x$-space. The parametric method with a rigid model advocated by the authors of Ref.~\cite{Chen:2025cxr} leads them to often not allow significant negative excursions, as for instance in both cases of Fig.~\ref{fig:fig1} and \ref{fig:fig2} although the data hints at a possible significant negative domain within uncertainty. This represents a modeling bias with no apparent physical justification.

Let us now consider unphysical PDF models which exhibit a sharp ``bump'' at $x = 0.5$. They are displayed in the lower-right plot of Fig.~\ref{fig:force_decay} with a dotted curve. Besides the bump, they are very similar to the model we have studied before. The bump in $x$-space translates into oscillations far in the Fourier domain. We introduce such unphysical models to study the repeated claim that LaMET is able to reconstruct point-like features in $x$-space, unlike other formalisms on the lattice. 

When we enforce the exponential decay in a similar fashion as above, the oscillations at large $\lambda$ are strongly reduced. As a result, the bump is completely erased when performing the inverse Fourier transform of the exponentially decaying models. This will of course not come as a surprise: if high frequencies are exponentially reduced with a decay length of $P_z / m_{\textrm{eff}}$, any sharp feature in $x$-space is smeared by a kernel with a width in $x$ of the order of $m_{\textrm{eff}}/P_z$. 

Reconstructing a point-by-point light-cone PDF, as claimed in Ref.~\cite{Chen:2025cxr}, requires to reliably compute the large Fourier harmonics corresponding to long distances on the light-cone. We know phenomenologically that those harmonics decay with a power-law. In the LaMET formalism, the large Fourier harmonics correspond to long space-like distances, which decay exponentially. As a result, the features of the light-cone PDF are only reconstructed in the LaMET framework up to a smearing of the order of $m_{\textrm{eff}}/P_z$. As $P_z$ increases, the quality of the reconstruction increases as is well understood. 

If the parton distribution is smooth, such that its typical variations occur on a scale in $x$ larger than $m_{\textrm{eff}}/P_z$, then it is not much of an issue. But then one must note that the relevant regulator of the inverse problem, in LaMET as in the pseudo-PDF formalism, is the assumed smoothness of the parton distributions, and not the asymptotic behavior of space-like correlators.

If one refuses to see the smoothness as the physically relevant regulator, then the exponential decay of space-like correlators becomes a nuisance since it does not bring useful information on the long-distance light-cone Fourier harmonics. In either case, the claim that LaMET can reconstruct a point-by-point $x$-dependence of light-cone PDFs, unlike other formalisms on the lattice, seems dubious.

\section*{Acknowledgments}
This project was supported by the U.S.~Department of Energy, Office of Science, Contract \#DE-AC05-06OR23177, under which Jefferson Science Associates, LLC operates Jefferson Lab. This work has benefited from the collaboration enabled by the Quark-Gluon Tomography (QGT) Topical Collaboration, U.S.~DOE Award \mbox{\#DE-SC0023646}. CJM is supported in part by U.S.~DOE ECA \mbox{\#DE-SC0023047}.
This research was funded, in part (HD and SZ), by l’Agence Nationale de la Recherche (ANR), project ANR-23-CE31-0019. AR acknowledges support by U.S.~DOE Grant \mbox{\#DE-FG02-97ER41028}. 
KO was supported in part by U.S.~DOE Grant \mbox{\#DE-FG02-04ER41302} and would like to acknowledge the hospitality of the American Academy in Rome, where he spent part of his sabbatical. 
The research was conducted in part (DR) under the Laboratory-Directed
Research and Development Program at Thomas Jefferson National Accelerator Facility for the U.S. Department of Energy.
Computations for this work were carried out in part on facilities of the USQCD Collaboration, which are funded by the Office of Science of the U.S.~Department of Energy. This work was performed in part using computing facilities at William \& Mary, which were provided by contributions from the National Science Foundation (MRI grant PHY-1626177), and the Commonwealth of Virginia Equipment Trust Fund. In addition, this work used resources at NERSC, a DOE Office of Science User Facility supported by the Office of Science of the U.S. Department of Energy under Contract \#DE-AC02-05CH11231, as well as resources of the Oak Ridge Leadership Computing Facility at the Oak Ridge National Laboratory, which is supported by the Office of Science of the U.S. Department of Energy under Contract No. \mbox{\#DE-AC05-00OR22725}.The authors acknowledge support as well as computing and storage resources by GENCI on Adastra (CINES), Jean-Zay (IDRIS) under project (2020-2024)-A0080511504.
The software codes {\tt Chroma} \cite{Edwards:2004sx}, {\tt QUDA} \cite{Clark:2009wm, Babich:2010mu}, {\tt QPhiX} \cite{QPhiX2}, and {\tt Redstar} \cite{Chen:2023zyy} were used in our work. The authors acknowledge support from the U.S. Department of Energy, Office of Science, Office of Advanced Scientific Computing Research, and Office of Nuclear Physics, Scientific Discovery through Advanced Computing (SciDAC) program, and from the U.S. Department of Energy Exascale Computing Project (ECP). The authors also acknowledge the Texas Advanced Computing Center (TACC) at The University of Texas at Austin for providing HPC resources, like the Frontera computing system~\cite{frontera}, which has contributed to the research results reported within this paper. The authors acknowledge William \& Mary Research Computing for providing computational resources and/or technical support that have contributed to the results reported within this paper.

\bibliography{biblio.bib}

\end{document}